\documentstyle[11pt,newpasp, twoside, epsf]{article}

\begin{document}

\title{Structure and Evolution of Nearby UV-bright Starburst Galaxies}

\author{J.S. Gallagher, C.J. Conselice, and N. Homeier}
\affil{Department of Astronomy, University of Wisconsin, Madison, WI 53706}

\author{The WFPC2 Investigation Definition Team}

\begin{abstract}
Nearby (D$ \leq$ 100~Mpc) 
luminous blue starburst galaxies frequently show morphological 
evidence for recent involvement in mild collisions, or minor mergers 
where the disk survives. As a consequence UV-bright starbursts 
are preferentially seen in near face-on galaxies, and the postburst 
systems may become star-forming late-type galaxies. If this starburst 
evolutionary channel is important at moderate redshifts, then descendants 
of the faint blue galaxies could be the very common Sm-Sd field galaxies. 
\end{abstract}


\section{Objectives}

Starburst galaxies provide unique opportunities to study rapid 
phases of galactic evolution.
The Wisconsin starburst 
project focuses on nearby ultraviolet-bright 
starbursts selected from the 
spectrophotometric sample of Gallagher et al. (1989).  They resemble 
intermediate redshift ($z \approx$ 0.5-1) ``compact narrow emission line 
galaxies'' (CNELGs, e.g. Guzeman et al. 1998) in terms of 
their small sizes, blue luminosities, 
and [OII] emission line equivalent widths; e.g., 
M$_B < -$18, and EW([OII]) $>$ 25 \AA\ with $U-B< -$0.2.

Our strategy is based on the analysis of 
multi-band imaging and area spectroscopy 
obtained  with the WIYN 3.5-m telescope \footnote{The WIYN 
Observatory is a joint facility of the University of
Wisconsin-Madison, Indiana University, Yale University, and the
National Optical Astronomy Observatories.} and high angular 
resolution WFPC2 and FOC {\it HST} images.  Images provide 
global morphologies, colors, and the small scale structures of 
star-forming sites. Spectra from the WIYN Densepak multi-fiber 
array are used to measure global velocity fields in 
the ionized gas (eg., Homeier \& Gallagher 1999; HG99).  

\section{Interactions and Starbursts}

Interactions are well-established starburst triggers.
Because of the huge range of possible 
interaction parameters, we expect and observe  
considerable variance in the responses of galaxies to interactions. 
Our emphasis is on weak interactions, such as minor mergers, or 
mild collisions, in which at 
least one member of the colliding pair survives with its initial structure 
relatively intact.

The signatures of recent 
galaxy interactions are well-known, and in the optical 
include: luminous tails 
of tidal debris, or ripples in post-interaction systems. 
These features often have low surface brightnesses and/or small 
angular sizes; imaging with telescopes like WIYN, that 
combine excellent surface brightness sensitivity with good 
angular resolution (0.7 arcsec seeing), is
valuable for determining the structures of these types 
of systems. The signposts of interactions evolve on galaxy orbital time scales 
of $\sim$10$^8$~yr with infall of tidal debris possibly 
continuing for $\sim$10$^9$~yr; after this, evidence of a collision 
will be difficult to detect.

Starbursts are found from their prominent populations  
of high mass stars, which are the products of upward 
spikes in star formation rates. This is obvious  
when the starburst is optically visible, and gives rise to high surface 
brightnesses, blue optical colors, and strong emission lines.
Starbursts also produce supergiant HII complexes, massive and 
compact ``super star clusters'', and the peculiar kpc-scale 
``clumps'' of OB stars, whose existence was 
emphasized more than 20 years ago by J. Heidmann and collaborators 
as symptoms of `hyperactive star formation'. While the 
evolutionary time scales of individual star-forming complexes within 
starbursts are not well known, data from M82 suggest this occurs  
in less than 10$^8$~yr (Satyapal et al. 1997; Gallagher \& Smith 1999).
The internal structure of a starburst evolves 
more quickly than structural perturbations induced by 
the dynamical influence of a colliding galaxy.

\section{Four Examples of Luminous Blue Starbursts}

The optical structures 
of four luminous starbursts within 100~Mpc provide some examples of 
star formation produced in a variety of collisions:

{\bf Markarian 8} is a strongly interacting system (Figure 1). It apparently 
consists of two small disk galaxies. Our WFPC2 
images show one of these to be strongly disturbed, while the other 
remains an organized, coplanar system  
(Gallagher et al. 2000). Both objects have high inclinations, but  
the system still has blue colors and strong [OII] emission.

\begin{figure}
\plottwo{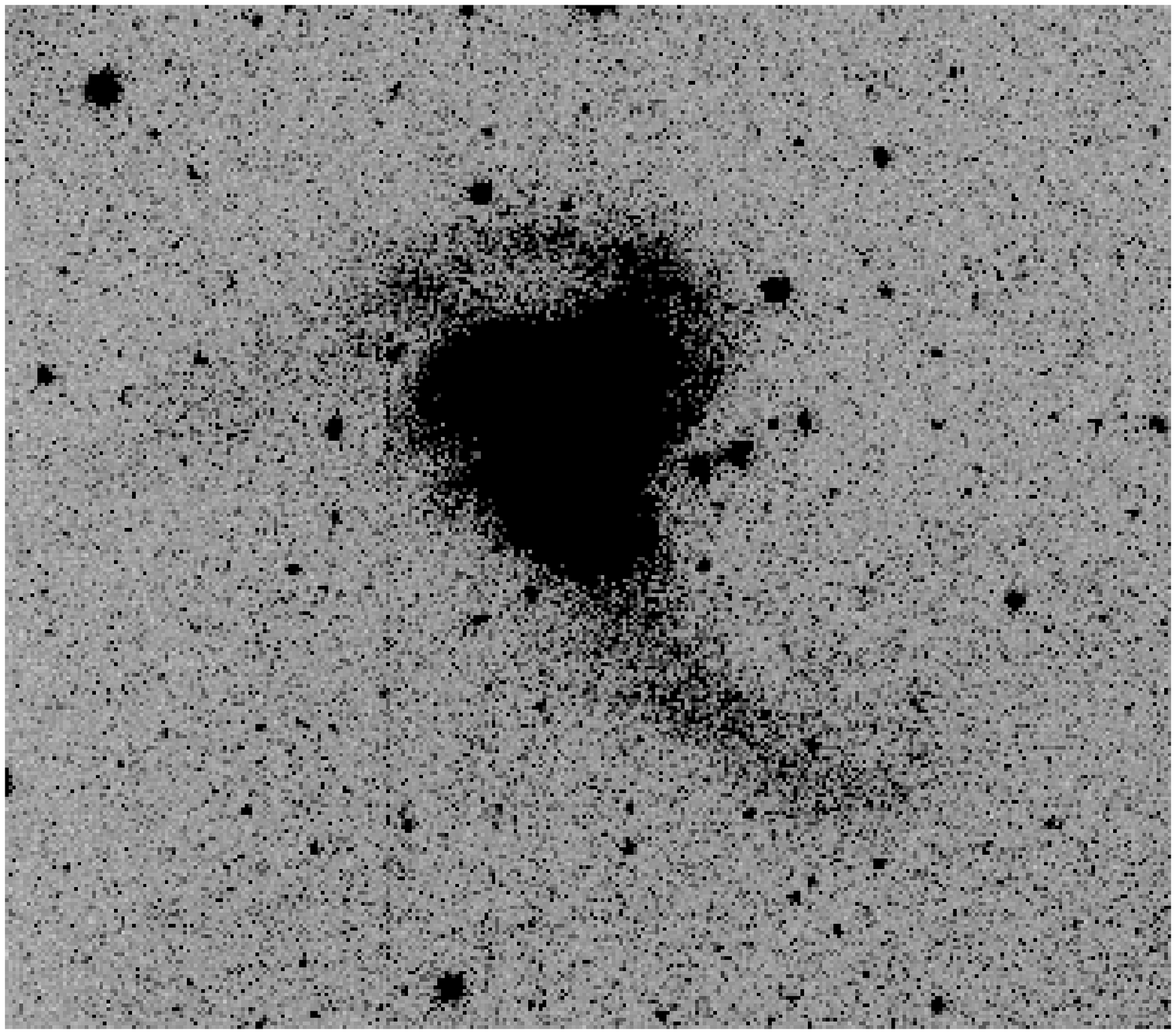}{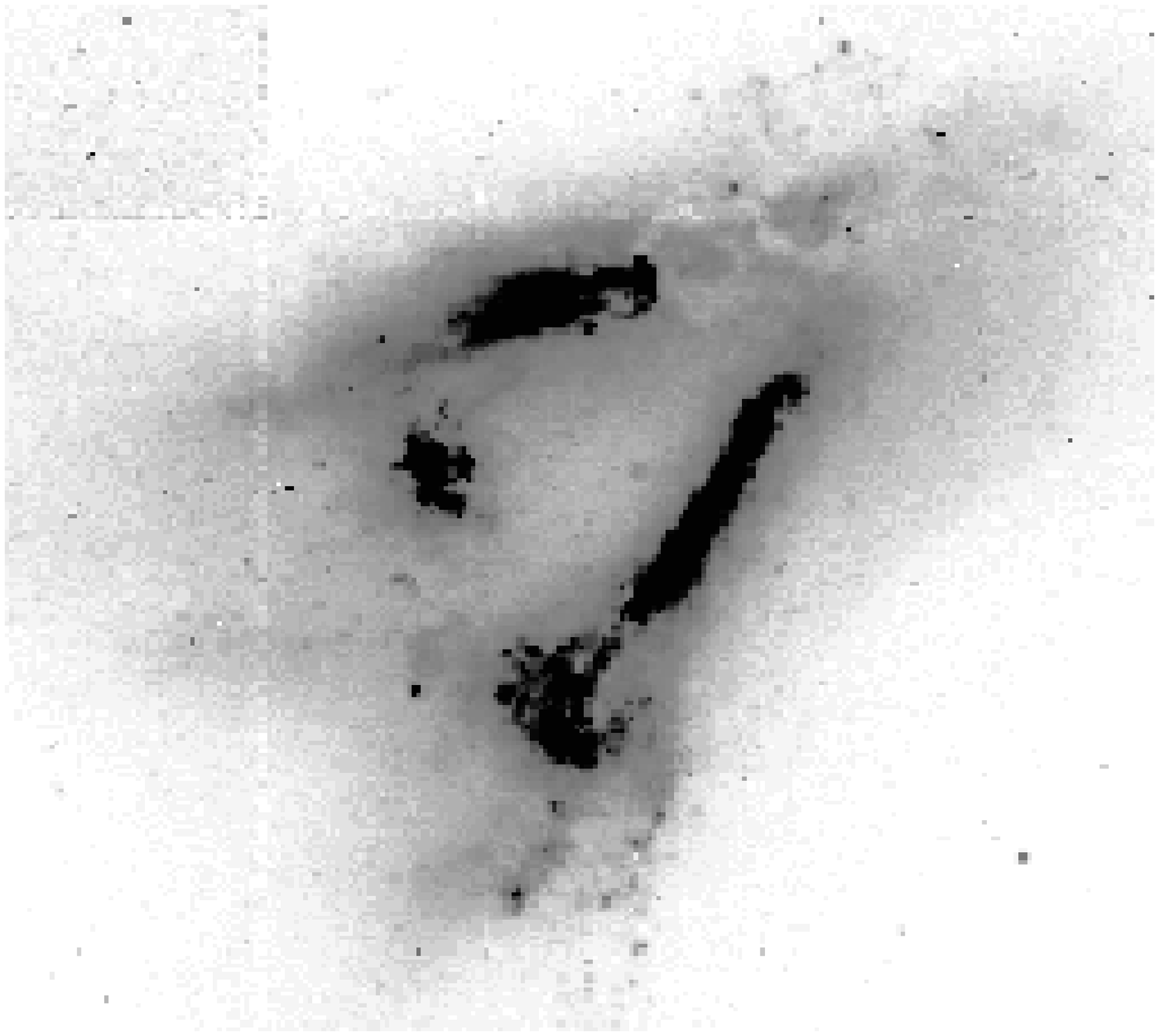}
\caption{\footnotesize{The Markarian 8 pair of colliding galaxies. View of the 
tidal tails from a WIYN R-band image (left) and high resolution image of the 
center from WFPC2.}}
\end{figure}

{\bf Haro 1}, a nearly face-on galaxy,  can be seen in Figure 2 to be 
experiencing a weak interaction 
with a dwarf companion. Even though only mild optical distortions are present   
in Haro 1, it is in a late phase of an intense 
starburst (the integrated spectrum shows strong Balmer absorptions). 

{\bf NGC 3310} is one of the nearest examples of a luminous UV-bright 
starburst and has been extensively studied in the optical and IR 
and mapped in HI. The starburst probably 
was caused by a minor merger (Mulder et al. 1995, Conselice et al. 2000). 
Large scale organization of star formation in NGC~3310 is located 
in a nearly symmetric two-arm spiral pattern, 
and around a mildly distorted circum-nuclear ring, as shown in the color map 
in Figure 2.  Evidently a strong starburst was initiated with 
little collisional disruption to the surviving gas-rich disk galaxy, 
seen at low inclination.

\begin{figure}[ht]
\plottwo{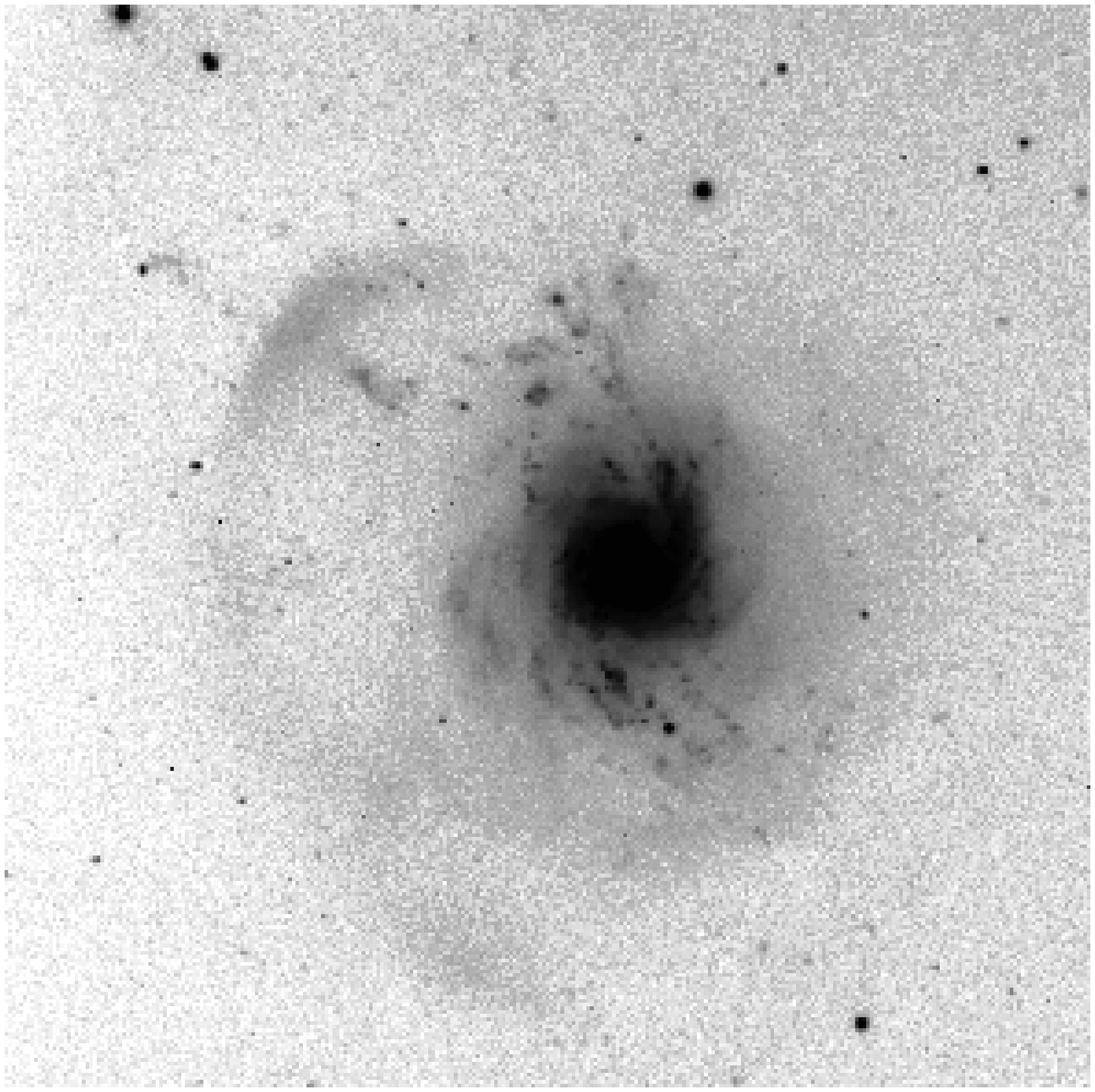}{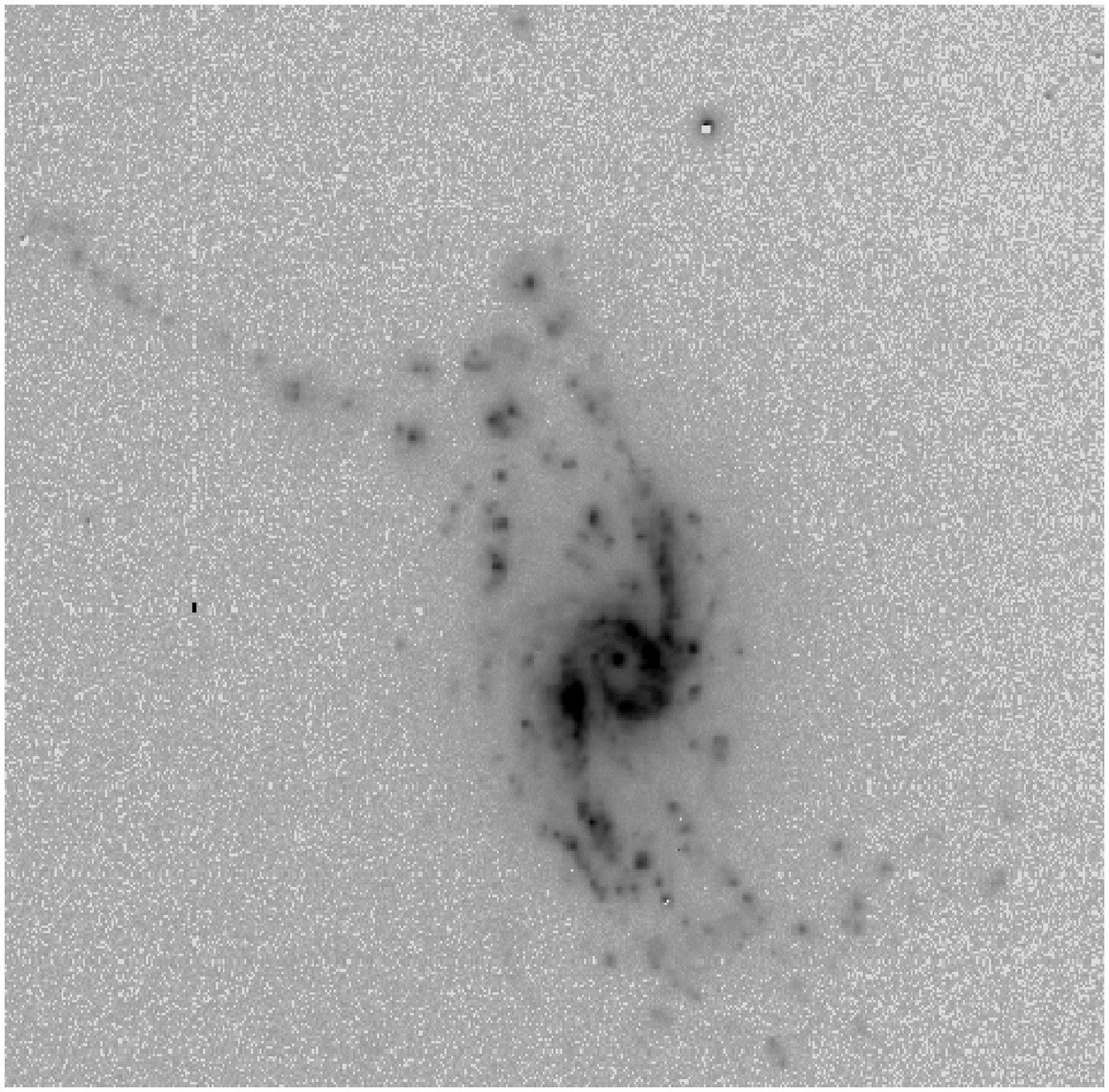}

\plottwo{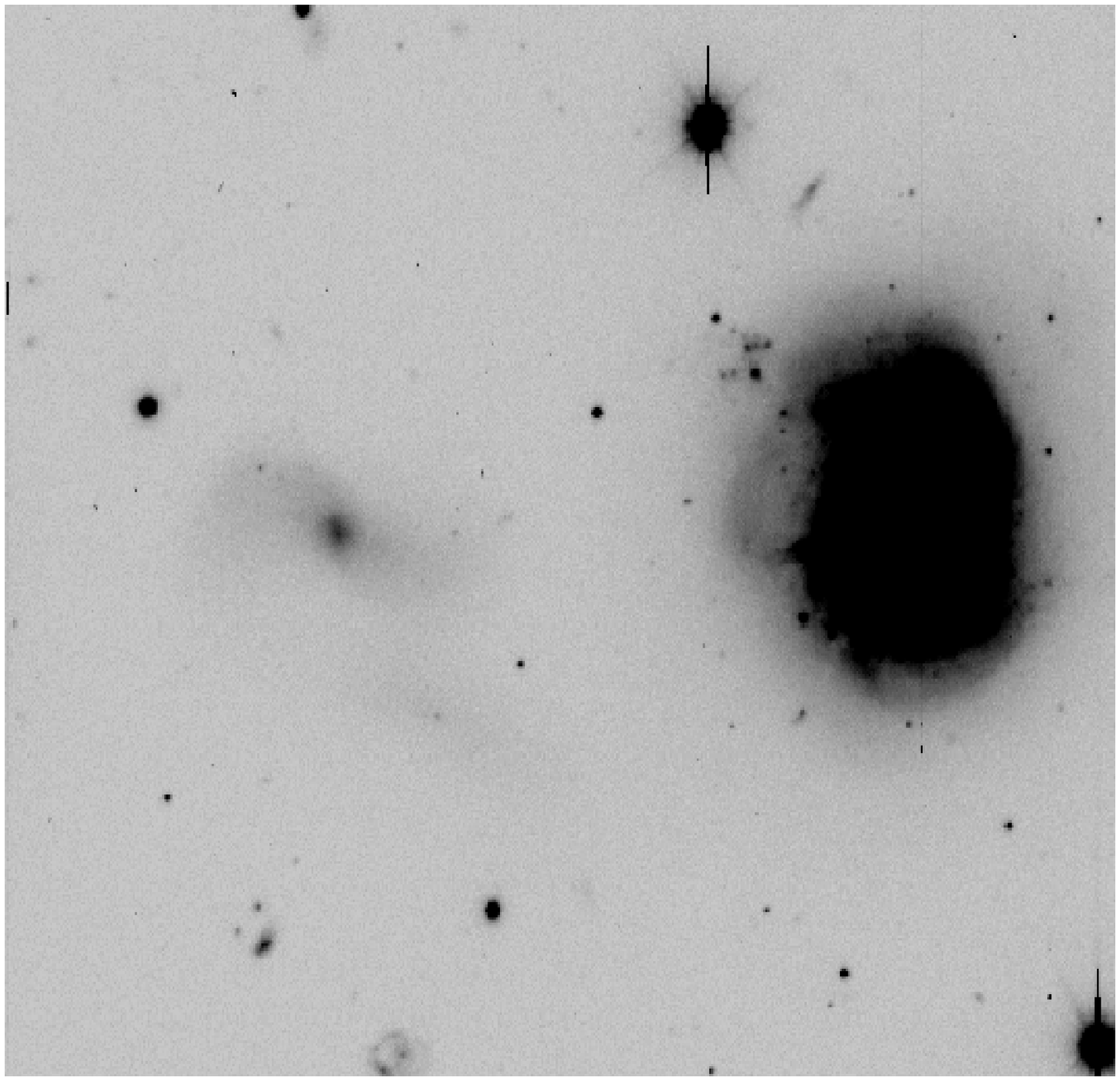}{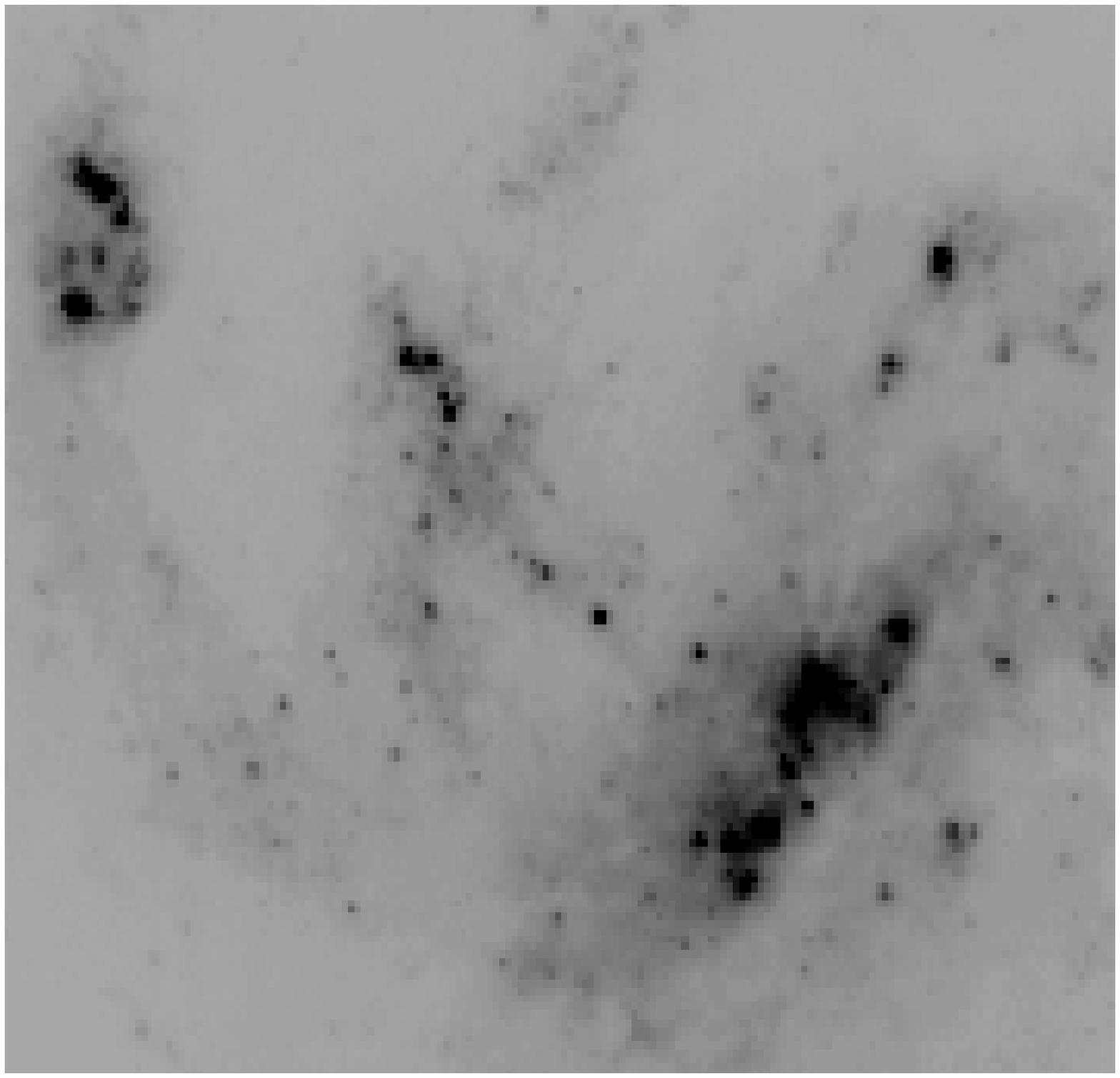}
\caption{\footnotesize{NGC 3310 WIYN R-band image (upper left), showing  
outer disk ripples and the `bow and arrow', left above the galaxy. 
The WIYN image in  
H$\alpha +$ [NII] emission lines (upper right) displays the nuclear ring 
and star formation along the spiral arms and arrow.  
A WIYN R-band view of Haro 1 (burned out) and its  
distorted dwarf companion is in the lower left. A WFPC2 F555W image of 
NGC~7673 showing resolved giant star-forming clumps (lower right).}}
\end{figure}

{\bf NGC 7673}, an archetypical `clumpy irregular', has a highly 
disturbed inner structure dominated by four huge star-forming 
complexes (Figure 2). WFPC2 optical and UV imaging resolve the clumps 
into clusters of compact star clusters, similar to conditions in 
M82 (O'Connell et al. 1995). The clumps are located within a faint, 
relatively symmetric outer disk that also contains a faint stellar 
arc, or ripple (HG99).  This starburst is a  
remnant of a past interaction with NGC~7678 (Nordgren et al. 1997), 
or a minor merger.

HG99 measured the H$\alpha$ velocity field of NGC~7673 with WIYN using 
the Densepak array. They found that the inner part of the galaxy is 
almost exactly face-on and dynamically cold. The wings of the 
line are broadened; most likely by gas outflows from the disk, and 
the integrated H$\alpha$ line profile closely resembles those observed 
in CNELGs by Guzman et al. (1996). However, in this case we know 
that we are {\it not} seeing a spheroidal system in formation, but 
rather a nearly face-on disk.

\section{Summary of Observed  Properties}

Our data demonstrate the production of intense starbursts 
in weak interactions in which at least one of the 
original disks survives. One channel for the production of 
compact luminous galaxies therefore modifies rather than 
destroys stellar disks, and some post-starburst 
systems will be small disk galaxies.

Disk geometries introduce a prejudice for optically- or UV- bright starbursts 
to be seen at low inclinations where disk obscuration is minimized. 
M82 is a nearby example; from our perspective it is an IR-starburst, 
but the brilliant UV reflection nebula seen above the disk in FOCAS 
images imply that when seen at low inclination, M82 is a UV-bright 
starburst. This is a statistical 
trend; some galaxies, such as Markarian 8 or chain galaxies at moderate 
redshifts, are blue and inclined. 
A bias to select blue starbursts in low inclination disks 
will also lead to narrow velocity line widths in integrated optical spectra; 
these are not necessarily signatures of low mass dwarf galaxies.

Our data do not show a simple relationship between the stage of an interaction 
and starburst properties; intense starbursts occur in both the ongoing 
Markarian 8 and very late NGC~7673 interactions. This complicates our 
ability to distinguish collisionally induced starbursts, since the 
clear signatures of an ongoing interaction may not be present during the 
peak starburst phase. This problem is especially severe when dealing with 
faint blue galaxies at moderate redshifts, where information is often 
sparse (Ellis 1997).

\section{Discussion}

A simple model assumes 
that luminous blue starbursts take place 
in gas-rich late-type disk galaxies that remain as disks after the event. 
Many nearby starbursts are rich 
in HI and have the fuel to support post-burst star-formation. 
These types of post-burst objects would be less easily distinguished 
than extreme cases where star formation is truncated after a burst,  
giving rise to distinctive `E $+$ A' 
spectra. The ingredients of our simple recipe for 
luminous starbursts follows:

1. Store gas in a dynamically cold, low density disk. Such disks are 
excellent storage places to keep gas as they have low intrinsic rates 
of evolution, and populations of such thin, cold disk galaxies exist 
at the current epoch (Matthews et al. 2000).

2. Perturb the disk with an interaction; a merger with a 
galaxy having 10\% or less 
of the total mass, or even a near miss from a comparable or larger 
neighbor should suffice.  What we require is that gas be 
driven to the central few kpc of the disk where it can support 
intense star formation. Gas transport can be produced directly by the 
perturber, or indirectly through production of a bar.

3. Orient galaxy to be nearly face-on and observe a blue, UV-bright 
starburst with narrow emission lines. Otherwise, observe an edge-on 
system to see an M82-type moderately-luminous IR-starburst.

4. Wait a few Gyr until done. The result may be a 
late-type system with a moderately thick stellar disk, a product of  
collisional disk heating during the interaction (Reshetnikov \& 
Combes 1997). The interaction also could produce an off-center bar, 
yielding a Magellanic irregular. This channel for making 
Magellanic systems is observationally suggested by the preference 
for such galaxies to have companions (Odewahn 1994) and theoretical 
models that produce characteristic off-center bars during collisions 
(Levine \& Sparke 1998).
\acknowledgments

The Wisconsin exploration of starburst galaxies has been supported as 
part of the WFPC2 Investigation Team's research program. We are also 
pleased to acknowledge support through General Observer Space Telescope 
grant, and from a Vilas Associateship award to JSG.
 
\footnotesize{

}

\end{document}